\documentclass{emulateapj}

\shorttitle{MODE IDENTIFICATION IN {\it KEPLER} F STARS}
\shortauthors{WHITE ET AL.}

\begin{document}

\title{Solving the mode identification problem in asteroseismology of F stars observed with {\it Kepler}}

\author{
Timothy~R.~White\altaffilmark{1,2}, 
Timothy~R.~Bedding\altaffilmark{1}, 
Michael~Gruberbauer\altaffilmark{3}, 
Othman~Benomar\altaffilmark{1}, 
Dennis~Stello\altaffilmark{1}, 
Thierry~Appourchaux\altaffilmark{4}, 
William~J.~Chaplin\altaffilmark{5}, 
J{\o}rgen~Christensen-Dalsgaard\altaffilmark{6}, 
Yvonne~P.~Elsworth\altaffilmark{5}, 
Rafael~A.~Garc\'\i a\altaffilmark{7}, 
Saskia~Hekker\altaffilmark{8,5}, 
Daniel~Huber\altaffilmark{1,9}, 
Hans~Kjeldsen\altaffilmark{6},  
Beno\^it~Mosser\altaffilmark{10},
Karen~Kinemuchi\altaffilmark{11},
Fergal~Mullally\altaffilmark{12}, and
Martin~Still\altaffilmark{11}
}
\affil{\altaffilmark{1}Sydney Institute for Astronomy (SIfA), School of Physics, University of Sydney, NSW 2006, Australia; t.white@physics.usyd.edu.au}
\affil{\altaffilmark{2}Australian Astronomical Observatory, PO Box 296, Epping NSW 1710, Australia}
\affil{\altaffilmark{3}Institute for Computational Astrophysics, Department of Astronomy and Physics, Saint Mary's University, Halifax, NS B3H 3C3, Canada}
\affil{\altaffilmark{4}Universit\'e Paris-Sud, Institut d’Astrophysique Spatiale, UMR8617, CNRS, Batiment 121, 91405 Orsay Cedex, France}
\affil{\altaffilmark{5}School of Physics and Astronomy, University of Birmingham, Birmingham B15 2TT, UK}
\affil{\altaffilmark{6}Danish AsteroSeismology Centre (DASC), Department of Physics and Astronomy, Aarhus University, DK-8000 Aarhus C, Denmark}
\affil{\altaffilmark{7}Laboratoire AIM, CEA/DSM-CNRS, Universit\'e Paris 7 Diderot, IRFU/SAp, Centre de Saclay, 91191, Gif-sur-Yvette, France}
\affil{\altaffilmark{8}Astronomical Institute `Anton Pannekoek', University of Amsterdam, Science Park 904, 1098 XH Amsterdam, The Netherlands}
\affil{\altaffilmark{9}NASA Ames Research Center, Moffett Field, CA 94035, USA}
\affil{\altaffilmark{10}LESIA, CNRS, Universit\'e Pierre et Marie Curie, Universit\'e Denis Diderot, Observatoire de Paris, 92195 Meudon cedex, France}
\affil{\altaffilmark{11}Bay Area Environmental Research Inst./NASA Ames Research Center, Moffett Field, CA 94035, USA}
\affil{\altaffilmark{12}SETI Institute/NASA Ames Research Center, Moffett Field, CA 94035, USA}

\begin{abstract}
\noindent 
Asteroseismology of F-type stars has been hindered by an ambiguity in identification of their oscillation modes. 
The regular mode pattern that makes this task trivial in cooler stars is masked by increased linewidths. 
The absolute mode frequencies, encapsulated in the asteroseismic variable $\epsilon$, can help solve this impasse because 
the values of $\epsilon$ implied by the two possible mode identifications are distinct. We find that the correct 
$\epsilon$ can be deduced from the effective temperature and the linewidths and we apply these methods to a 
sample of solar-like oscillators observed with {\it Kepler}.
\end{abstract}


\section{Introduction}
Asteroseismology of solar-like stars is developing rapidly, driven by the successes of the space telescopes CoRoT 
\citep{Michel08} and {\it Kepler} \citep{Koch10,Gilliland10,Chaplin11a}. By studying the oscillation modes of these 
stars, inferences can be made about their interior structures \citep[e.g.][]{Verner11}. Except for the most basic analyses, it is crucial 
to identify the oscillation modes, that is, the radial order $n$ and the spherical degree $l$.

In the Sun and similar stars, mode identification is straightforward thanks to the distinctive pattern of alternating
odd and even modes in the power spectrum. This pattern consists of a regular sequence of $l=1$ modes, interspersed with
close pairs of $l=0$ and 2 modes that fall approximately halfway in between. However, stars significantly hotter
than the Sun have large linewidths that blur the $l=0,2$ pairs and make mode identification very difficult. In this
Letter we demonstrate a solution to this problem that applies the method proposed by \citet{Bedding10b} and 
\citet{White11b}, which uses the {\em absolute} frequencies of the oscillation modes.

\section{Methods}

For main-sequence stars, the frequencies are well-approximated by the asymptotic 
relation \citep{Vandakurov67, Tassoul80, Gough86},
\begin{equation}
\nu_{n,l}\approx\Delta\nu\left(n+{l \over 2}+\epsilon\right)-\delta\nu_{0l}.\label{asymp}
\end{equation}
Here, $\Delta\nu$ is the large separation between modes of the same degree $l$ and consecutive order $n$, $\delta\nu_{0l}$ is
the small separation between modes of different degree and $\epsilon$ is a dimensionless offset, which we discuss
in greater detail below. Typically, only modes of $l\le 2$ are observed in intensity due to cancellation over the unresolved
stellar disk, although $l=3$ modes can be observed in the highest signal-to-noise targets, 
such as 11 {\it Kepler} subgiants for which frequencies have been determined by \citet{Appourchaux12b} and the
solar analogs 16~Cyg~A~and~B \citet{Metcalfe12}.

\begin{figure*}
\epsscale{1.2}
\plotone{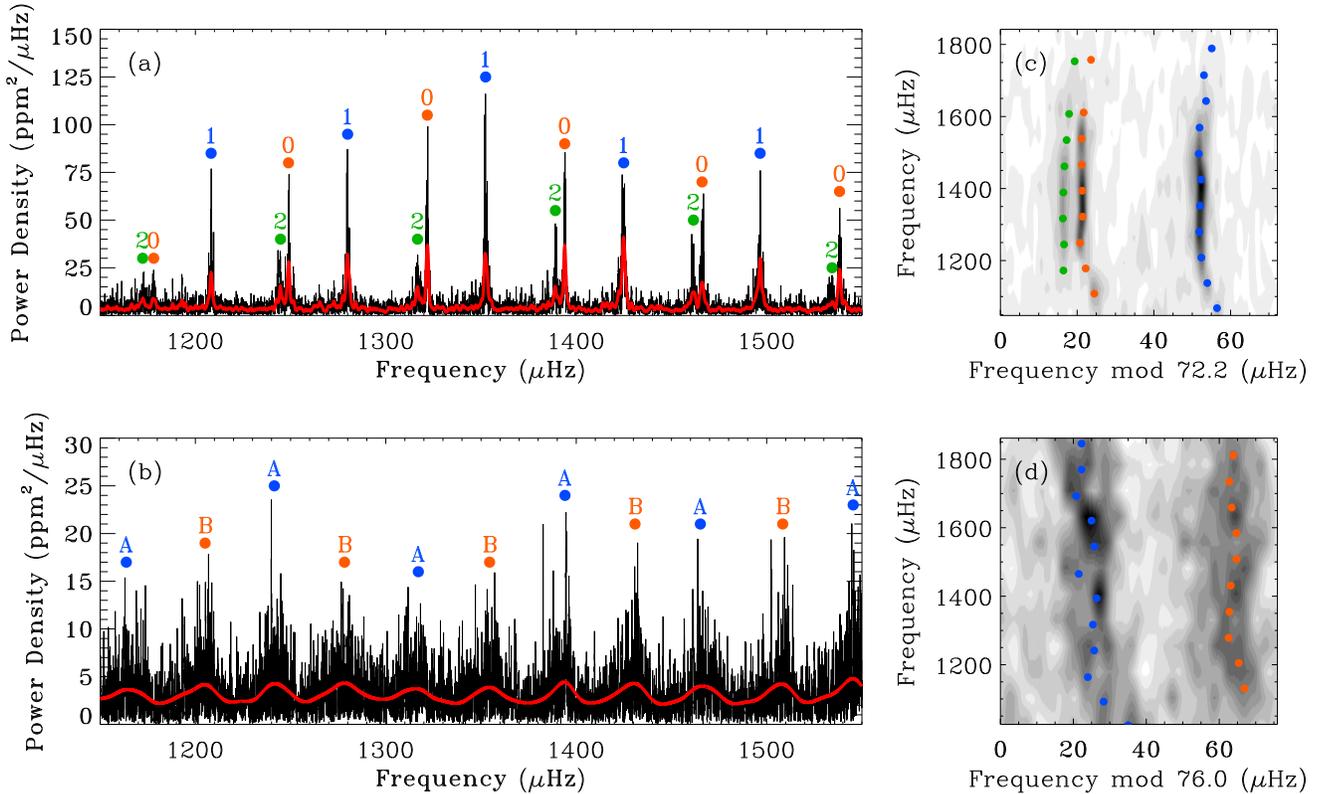}
\caption{Power spectra of (a) a G star, KIC 6933899, and (b) an F star, KIC 2837475, with their corresponding \'echelle
diagrams (c) and (d), respectively. The red curves show the power spectra after smoothing. Mode identification of the
G star is trivial, with modes of $l=0$ (orange), 1 (blue) and 2 (green) labeled. For the F star it is not clear
whether the peaks labeled `A' (blue) or `B' (orange) correspond to the $l=1$ or $l=0,2$ modes.}\label{fig1}
\end{figure*}

The asymptotic relation makes it easy to determine the mode degrees for the Sun and similar stars. Each
$l=0$ mode is separated by $\delta\nu_{02}$ from an $l=2$, and separated by $\Delta\nu/2-\delta\nu_{01}$ from the $l=1$ mode 
of the same order. An example is shown in Figure~\ref{fig1}a for the {\it Kepler} star KIC~6933899, 
which has an effective temperature of 5840\,K.

Mode linewidth increases with effective temperature \citep{Chaplin09,Baudin11,Appourchaux12a, Corsaro12}, reflecting
shorter mode lifetimes in hotter stars. In some F-type stars, the linewidths become so large that the pairs of $l=0$ and $l=2$ modes
are unresolved and it becomes difficult to distinguish between the blended $l=0,2$ modes and the $l=1$ modes. This
problem was first observed by CoRoT, in the F5 star HD\,49933 \citep{Appourchaux08} and has since been seen in other 
CoRoT stars \citep{Barban09,Garcia09} and in the bright F5 star Procyon \citep{Bedding10}. We also see it in many
{\it Kepler} stars and Figure~\ref{fig1}b shows one example, KIC~2837475 ($T_\mathrm{eff} = 6690$\,K).

One way to resolve this identification problem is to fit both possible mode identifications and compare 
the relative likelihoods of the two scenarios \citep{Appourchaux08,Benomar09,Gruberbauer09,Kallinger10,Bedding10,Handberg11}. 
This relies on the profile of the even-$l$ modes being significantly broader and also asymmetric, relative to the $l=1$ 
modes (owing to the presence of the smaller amplitude $l=2$ modes at a slightly lower frequency than the $l=0$ modes). 
The correct scenario should provide a better fit to the power spectrum. Difficulties 
arise at low signal-to-noise and with short observations, for which the Lorentzian mode profiles are not well 
resolved. This method was first applied by \citet{Appourchaux08}, who fitted both scenarios for HD\,49933. 
However, with additional data their preferred mode identification was overturned by \citet{Benomar09}.

Other methods have been suggested that utilize the sign of the small separation $\delta\nu_{01}$ \citep{Roxburgh09,Mosser09}. 
In main sequence stars like the Sun, $\delta\nu_{01}$ is known to be positive. However, in many red giants 
$\delta\nu_{01}$ is found to be negative \citep{Bedding10c,Huber10,Mosser11}, so at some point in the evolution the sign 
must flip \citep{Stello11}. To further complicate matters, the value of $\delta\nu_{01}$ is quite
small. At low signal-to-noise, it may be difficult to obtain frequencies precise enough to determine the sign
of $\delta\nu_{01}$ reliably.

\citet{Bedding10b} have suggested that scaling the frequencies of a star with a known mode identification could reveal 
the correct mode identification in a second star. This method seeks to use information contained within the value of 
$\epsilon$. For this to be effective, $\epsilon$ must vary slowly as a function of stellar parameters. 
This is indeed the case, with a tight relationship between $\epsilon$ 
and effective temperature, $T_\mathrm{eff}$, found both in models \citep{White11a} and in observations of Sun-like stars 
\citep{White11b}. Thus, $\epsilon$ promises to be an effective way to determine mode identifications, since the 
difference in the value of $\epsilon$ for the two possible scenarios is large (0.5).

The existence of a relation between $\epsilon$ and $T_\mathrm{eff}$ is not surprising. The value of $\epsilon$ 
is determined by the upper and lower turning points of the acoustic waves \citep[e.g.][]{Gough86}. As such, 
$\epsilon$ is heavily dependent upon the stellar atmosphere, of which $T_\mathrm{eff}$ is a significant parameter. 
Due to inadequate modeling of the near-surface layers, there is a well-known 
offset between observed and computed oscillation frequencies in the Sun \citep{C-D88b,Dziembowski88,C-D96,C-D97} and 
also in other stars \citep{Kjeldsen08,White11b,Mathur12}. This offset results in the computed $\epsilon$ being smaller 
than observed, typically by $\sim0.2$ as inferred from the displacement of model tracks from observations
in the $\epsilon$ diagram.

The purpose of this Letter is to extend the relationship between $\epsilon$ and $T_\mathrm{eff}$ to higher temperatures,
and thereby make reliable mode identifications in F-type stars.

\section{Observations and Data Analysis}
We used observations of solar-like oscillations in 163 stars taken with the {\it Kepler} space telescope
between May 2009 and March 2011 (Quarters 1--8). Each star was observed in {\it Kepler's} short-cadence mode 
(58.9 s sampling) for part of this period. The time series were prepared from the raw observations as described 
by \citet{Jenkins10} and further corrected to remove outliers and jumps as described by \citet{Garcia11}. 

\begin{figure}
\epsscale{1.2}
\plotone{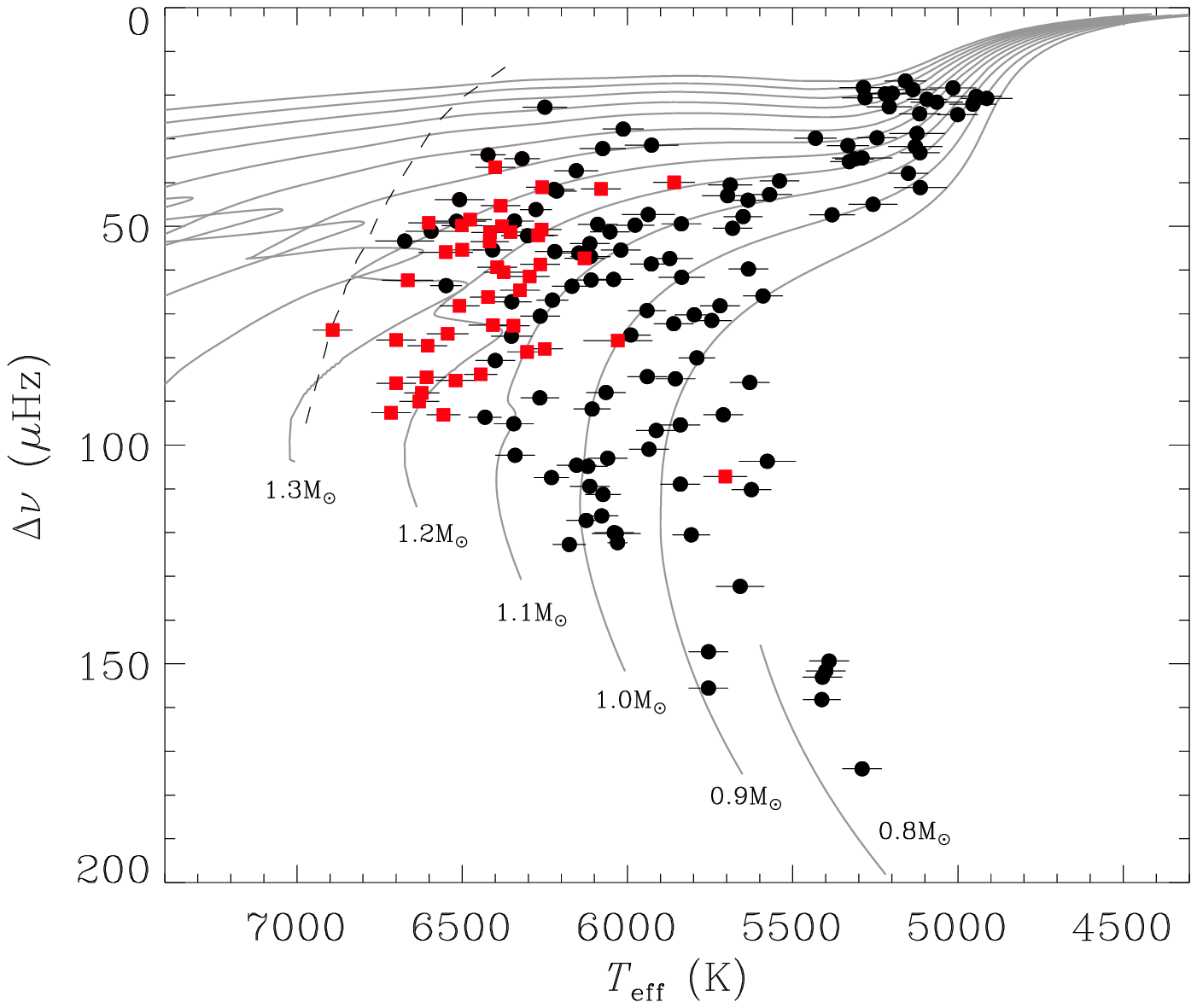}
\caption{Modified H-R diagram: average large frequency separation, $\Delta\nu$, against effective temperature for 
stars in our sample. Stars with secure mode identifications are indicated by black circles. Those without are
red squares. Grey lines are ASTEC \citep{C-D08a} evolutionary tracks for a metallicity of 
$Z_0 = 0.011$ ($[\mathrm{Fe/H}]= -0.2\,\mathrm{dex}$), matching the $T_\mathrm{eff}$ calibration of \citet{Pinsonneault11}. 
The dashed line indicates approximately the cool edge of the classical instability strip \citep{Saio98}. }
\label{fig2}
\end{figure}

Effective temperatures were determined from SDSS $griz$ color-temperature relations by \citet{Pinsonneault11}.
Spectroscopic temperatures have also been determined for 77~stars in our sample by \citet{Bruntt12}. In almost all
cases the photometric and spectroscopic temperatures agree, except for several stars where the
disagreement may be due to differing metallicities or unresolved binaries (the temperatures for
KIC\,3424541, 3456181, 4638884, 6679371, 7976303, 8938364, 9908400, 10018963, 10124866 and 10963065 were found to
disagree by more than 3$\sigma$). Figure~\ref{fig2} 
shows a modified HR diagram of this sample, where we have used large separation instead of luminosity.

To measure the value of $\epsilon$ for each star, we first determined the frequencies of the $l=0$ modes.
Where it was possible to resolve the $l=0$ and $l=2$ modes, we measured the frequencies of the $l=0$ modes
from the peak in the power spectrum after smoothing. This was possible for 115 stars. An example is shown in Figure 
\ref{fig1}a.

Where it was not possible to resolve the $l=0$ and $l=2$ modes, and therefore not possible to easily determine the correct 
mode identification, we determined $\epsilon$ for both scenarios (43 stars). In this case we used the frequencies of 
the ridge centroids, determined from the peaks of the heavily smoothed power spectrum, as shown in an 
example F star in Figure~\ref{fig1}b. There were also two cooler stars for which the mode 
identification was not clear (KIC\,11401708 and 12555505); these have low signal-to-noise and 
the $l=2$ modes are not apparent.

For five stars with blended $l=0$ and $l=2$ modes it was still possible to make an unambiguous mode identification
(KIC\,6064910, 6766513, 7668623, 7800289 and 8026226). 
In these stars, avoided crossings `bump' the $l=1$ modes from their asymptotically expected position 
\citep{Osaki75,Aizenman77}, revealing the correct identification. In these five cases, the 
frequencies were obtained from the centroids of the $l=0,2$ ridge.

The values of $\Delta\nu$ and $\epsilon$ were obtained from a weighted least-squares fit to the $l=0$ frequencies,
as described by \citet{White11a}. The weights were given by a Gaussian function centered at the
frequency of maximum power, $\nu_\mathrm{max}$, with a FWHM of $0.25\,\nu_\mathrm{max}$. By the asymptotic relation,
equation~\ref{asymp}, the gradient of this fit is $\Delta\nu$ and the y-intercept is $\epsilon\Delta\nu$.

To confirm the validity of our method of frequency determination, we compared the values of $\epsilon$ derived from 
our frequencies to those derived from frequencies that have been determined by more traditional `peak-bagging' methods
in the 61 stars for which this has been done \citep{Appourchaux12b}. We found good agreement, with 
$\epsilon$ values typically agreeing to within 0.1, which is approximately the size of the typical uncertainty. This 
agreement is of particular importance for hotter stars because using ridge centroids instead of $l=0$ frequencies 
potentially biases $\epsilon$ towards slightly lower values. We found this bias to be negligible within the uncertainties.

\section{Results}

\begin{figure*}
\epsscale{0.579}
\plotone{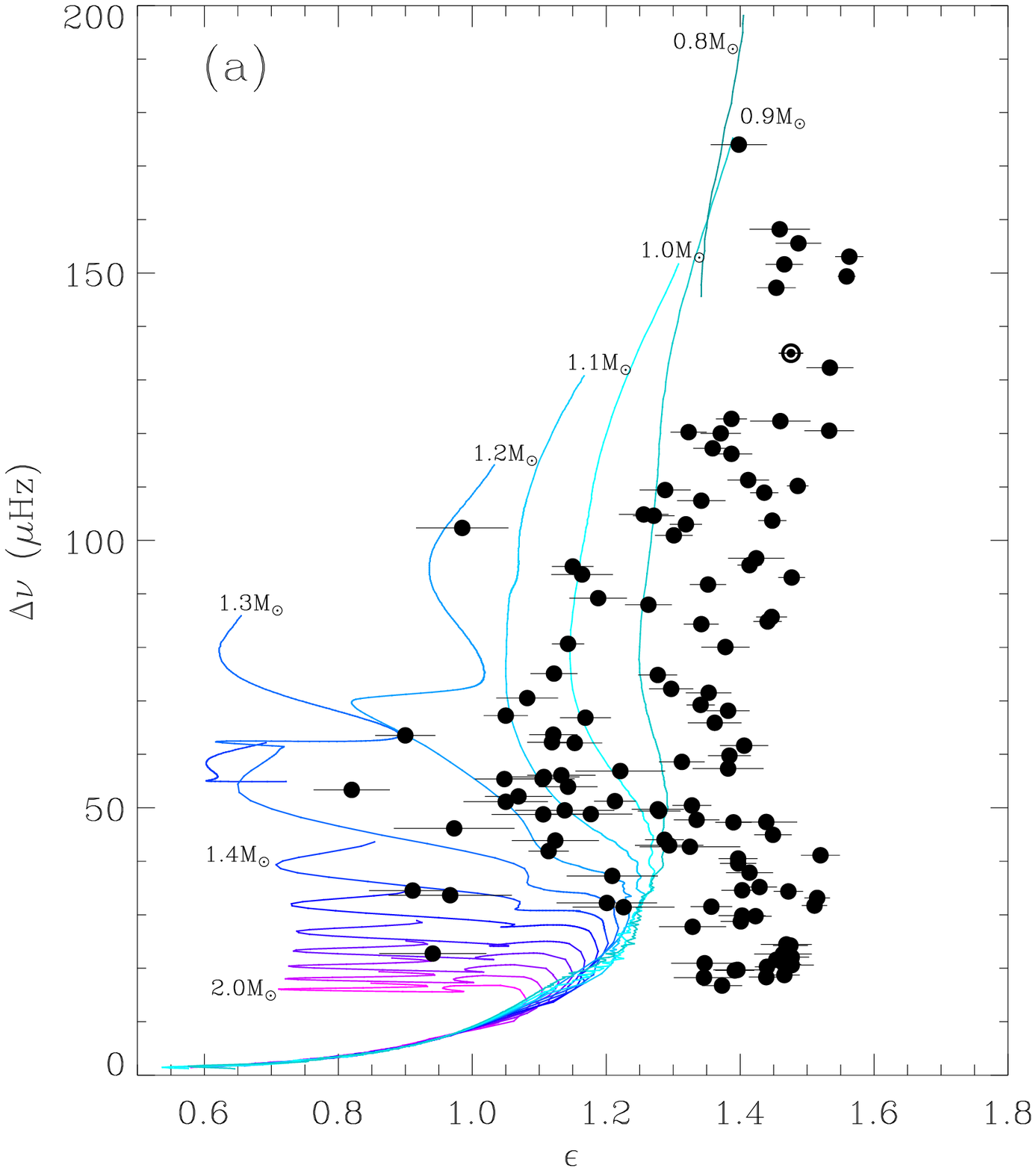}
\plotone{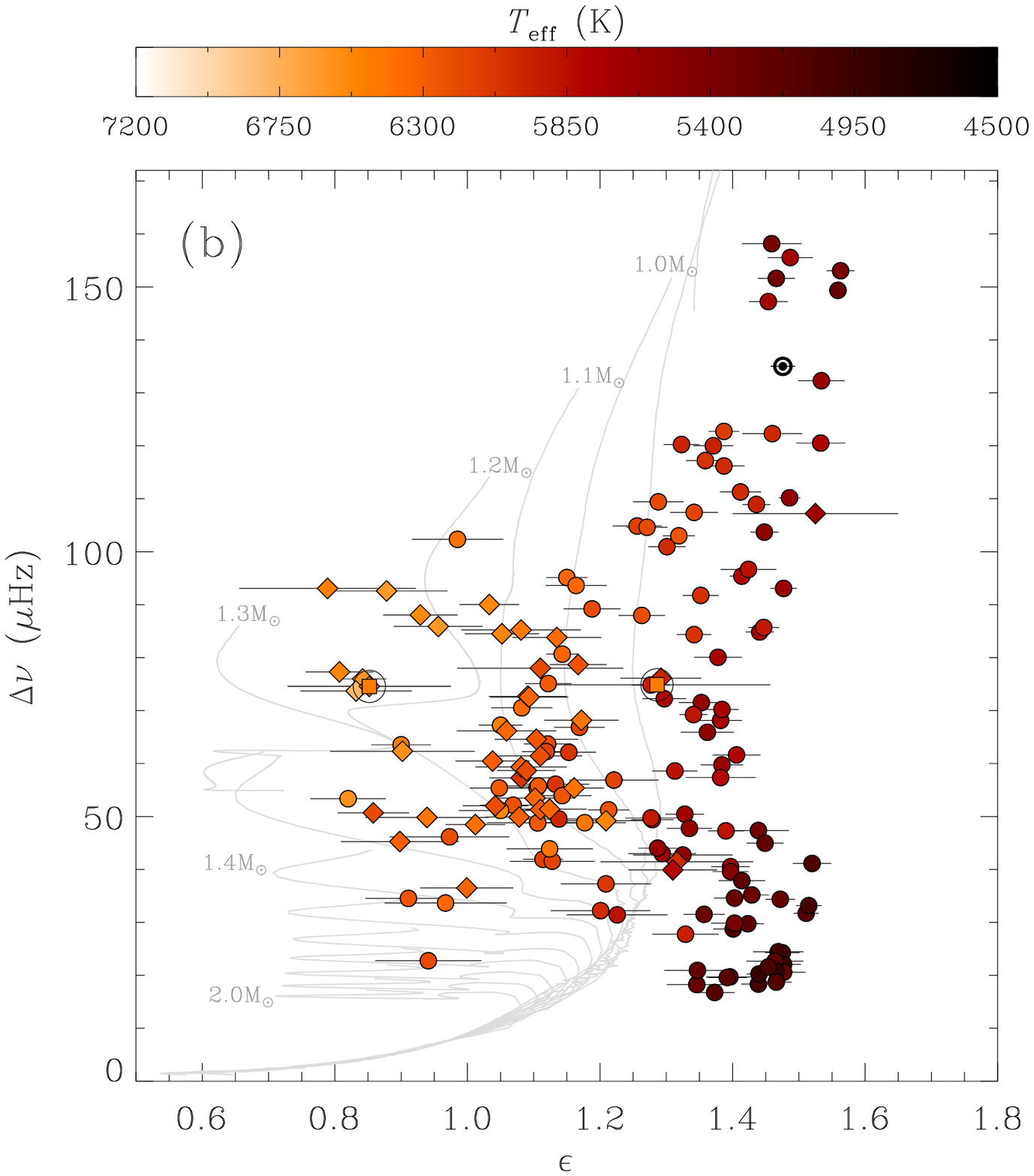}
\caption{The $\epsilon$ diagram: large separation, $\Delta\nu$, against $\epsilon$. (a) Only stars with secure
mode identifications are shown (filled black circles). Lines are ASTEC evolutionary tracks, as shown in Figure~\ref{fig2}, 
although for clarity, segments of the tracks which are hotter than the approximate cool edge of the instability strip
are not shown. Note the offset between models and observations. (b) Symbol colors reflect the measured
effective temperature of the star. Stars with obvious identifications are circles, and those for which we can
reliably make the identification from the $\epsilon$--$T_\mathrm{eff}$ relation are indicated by diamonds.
The possible identifications of one star for which the identification is still ambiguous in the 
$\epsilon$--$T_\mathrm{eff}$ plane is indicated by the encircled squares. Comparing the possible identifications 
with the temperatures of stars of similar $\Delta\nu$, the scenario on the left can be preferred.}\label{fig3}
\end{figure*}

In Figure~\ref{fig3}a we show the so-called $\epsilon$ diagram for the 120 {\it Kepler} stars whose mode 
identifications were unambiguous. The observed values of $\epsilon$ are clearly offset to the right of the models 
which, as mentioned above, arises from the improper modeling of the near-surface layers. From this figure it appears 
that the offset in $\epsilon$ may be roughly the same for all stars, which corresponds to a fixed fraction of the large 
separation, as was also inferred by \citet{Mathur12}.

\begin{figure}
\epsscale{1.1}
\plotone{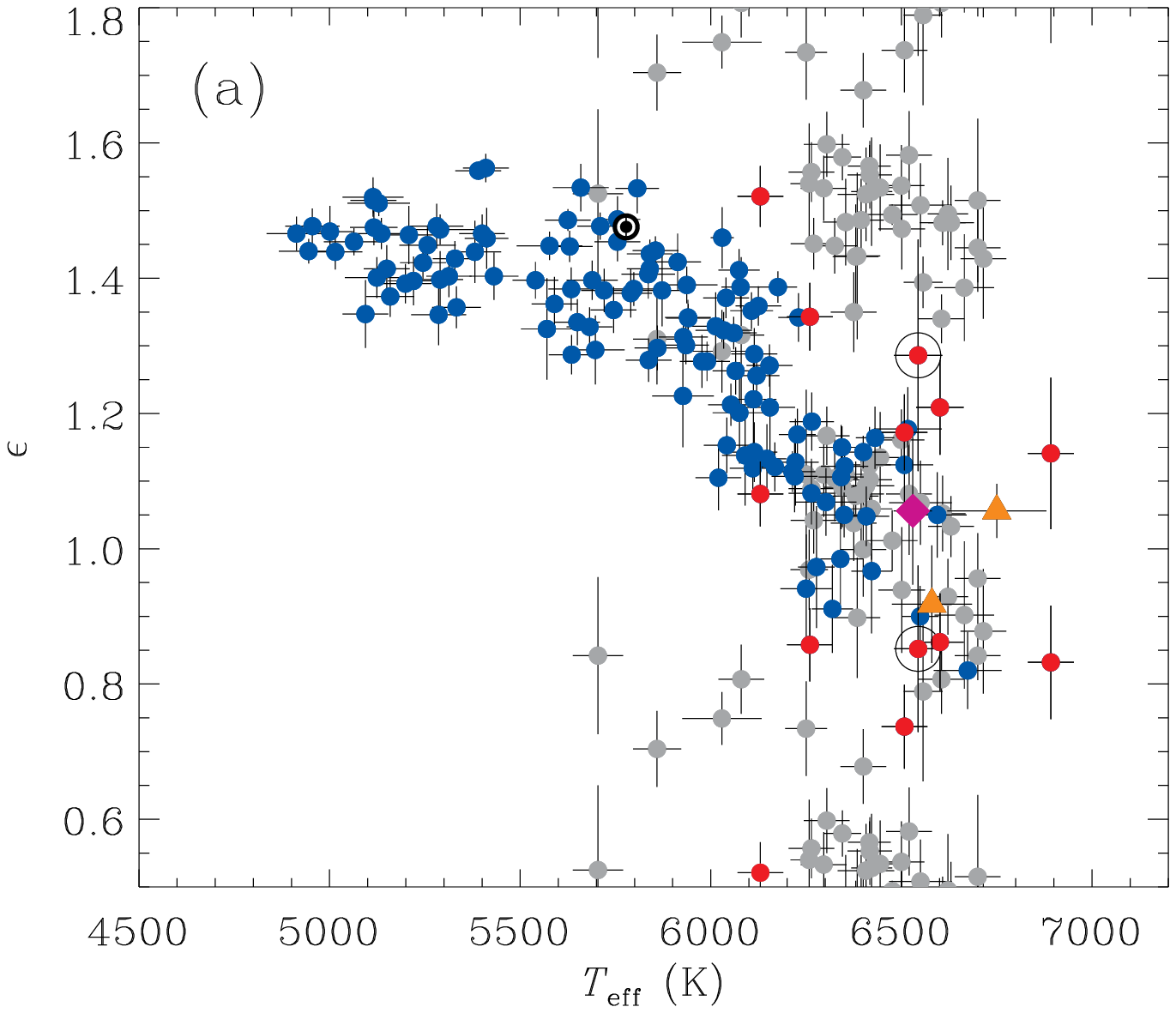}
\plotone{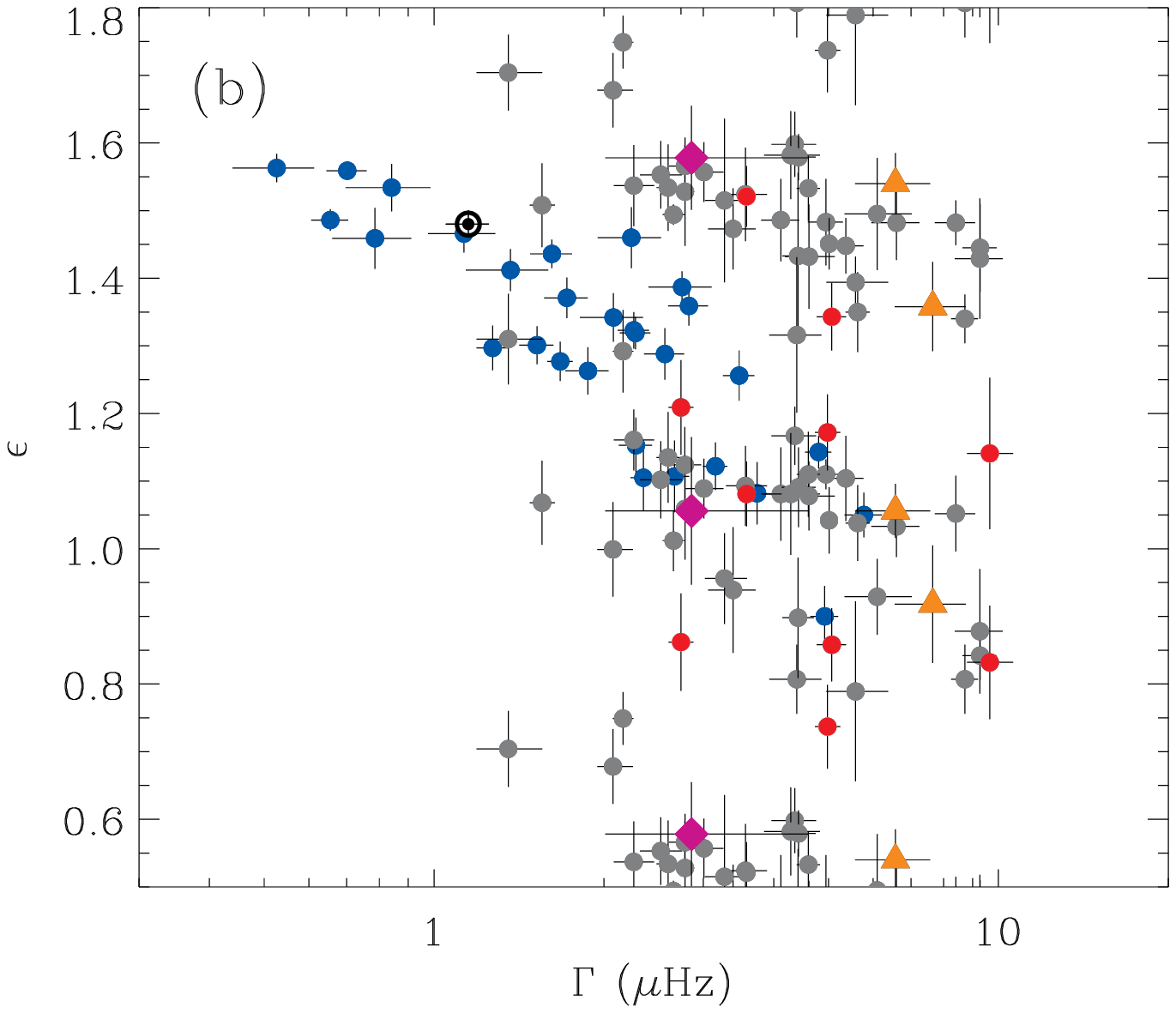}
\plotone{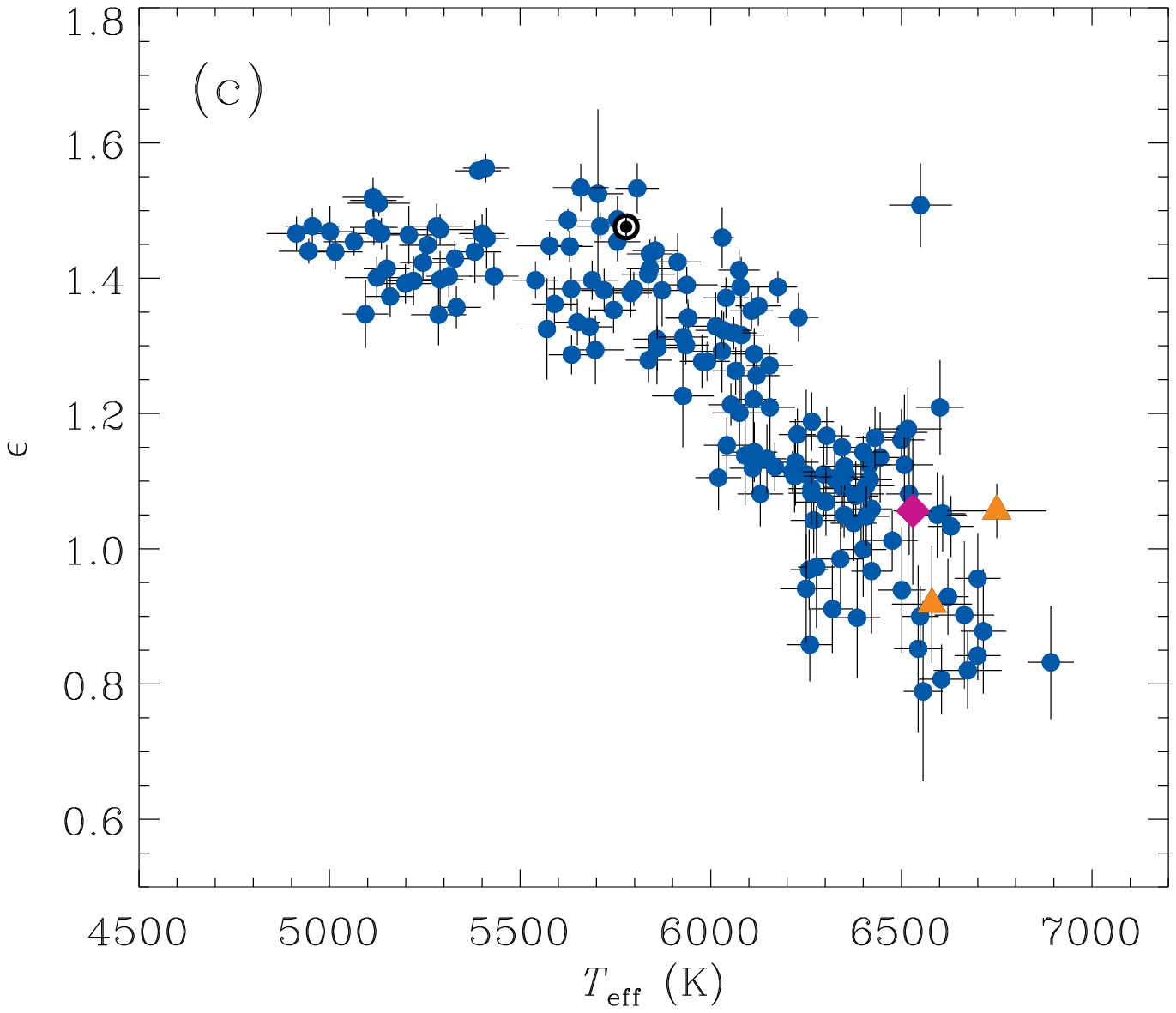}
\caption{(a) $\epsilon$ as a function of effective temperature. 
Stars with unambiguous mode identifications are indicated by blue circles. Stars with ambiguous identifications have 
two possible values of $\epsilon$ (gray circles) corresponding to the two possible identifications. Several stars 
for which the width of the relation between $\epsilon$ and $T_\mathrm{eff}$ makes the identification difficult are 
red circles. The scenarios of an example star for which we use $\Delta\nu$ to aid in the identification is circled, 
as it is in Figure~\ref{fig3}b. 
(b) Same as (a), but showing $\epsilon$ as a function of mode linewidth, $\Gamma$.
(c) The selected identification for all stars shown in the $\epsilon$--$T_\mathrm{eff}$ plane. The
outlier is KIC\,1725815, for which $\Gamma$ and $T_\mathrm{eff}^\mathrm{phot}$ disagree (see text).
In all panels the Sun is indicated by its usual symbol. Identifications of Procyon are indicated by magenta diamonds; 
those of the CoRoT F stars HD~49933 and HD~181420 are orange triangles.}
\label{fig4}
\end{figure}

In Figure~\ref{fig4}a we show $\epsilon$ versus $T_{\rm eff}$ for our sample. The Sun is marked in black
by its usual symbol, and the stars with unambiguous mode identifications are colored blue.  The trend of decreasing 
$\epsilon$ with increasing $T_\mathrm{eff}$ is clearly seen in these stars. 

For the 43 {\it Kepler} stars whose mode identifications are uncertain, the relationship between $\epsilon$ and 
$T_\mathrm{eff}$ can help. In Figure~\ref{fig4}a these stars are shown in gray for both scenarios. Owing to the potential
ambiguity in $n$, each scenario is also plotted shifted by $\pm1$. { We also include both scenarios of the F stars 
Procyon \citep{Bedding10}, HD~49933 \citep{Appourchaux08} and HD~181420 \citep{Barban09}.}

For most of the stars with ambiguous mode identifications, only one of the scenarios lies along the 
$\epsilon$--$T_\mathrm{eff}$ trend defined by the stars with secure identifications and we adopt this as
the correct one. { For the previously studied F stars, we prefer Scenario B of Procyon \citep{Bedding10}, Scenario B of
HD~49933 \citep{Benomar09} and Scenario 1 of HD~181420 \citep{Barban09}.} Due to the width of the 
$\epsilon$--$T_\mathrm{eff}$ relation, there are a few stars for which the situation is still somewhat ambiguous 
(red circles in Figure~\ref{fig4}a). The two scenarios in these stars have values of $\epsilon$ that fall towards the 
top and bottom of the relation. For these stars we must use additional information to resolve the ambiguity.

To overcome the spread in the $\epsilon$--$T_\mathrm{eff}$ relation, the value of $\Delta\nu$ is useful. 
For a given effective temperature, stars with higher masses have smaller values of $\Delta\nu$, as can be seen from 
the models in Figure~\ref{fig2}. Models also indicate that higher-mass stars have a smaller $\epsilon$ 
\citep[see Figure~\ref{fig3}a and][]{White11a}. It follows that in the $\epsilon$--$T_\mathrm{eff}$ plane, stars that 
fall towards the bottom of the trend will be  more massive and should therefore have smaller values of $\Delta\nu$ than 
lower-mass stars of similar temperature.

We illustrate this in the $\Delta\nu$--$\epsilon$ plane in Figure~\ref{fig3}b, where we show stars in which the 
identification was already obvious, as well as those for which the identification could be readily made from the 
$\epsilon$--$T_\mathrm{eff}$ relation. Symbols are colored according to $T_\mathrm{eff}$, with the trend of 
decreasing $T_\mathrm{eff}$ with increasing $\epsilon$ quite clear. The gradual decrease in $\epsilon$ with 
decreasing $\Delta\nu$ along lines of constant temperature is also visible. By comparing a star whose identification 
is still ambiguous with stars of a similar large separation, we can better make a decision on the identification. 
We show the example of KIC 11290197, whose two scenarios are circled in Figures~\ref{fig3}b~and~\ref{fig4}a. By 
comparing its $T_\mathrm{eff}$ with stars of a similar $\Delta\nu$, we establish the preferred scenario, which in this 
case has the lower value of $\epsilon$.

Other seismic parameters may also be useful. The method of \citet{Mosser10}, which uses the value of the
small separation $\delta\nu_{01}$, agrees for all stars, except one (KIC\,5431016), which has a low signal-to-noise.

Another very useful parameter is mode linewidth ($\Gamma$). As mentioned above, linewidth is 
strongly correlated with effective temperature, and so there should also be a correlation between linewidth 
and~$\epsilon$. Linewidths for~41 of our stars (including 12 with ambiguous identifications) were 
previously measured by \citet{Appourchaux12a}. Using the SYD method described in that paper, we have measured 
linewidths in a further~26 stars with ambiguous identifications (excluding the
five ambiguous stars with the lowest signal-to-noise). Figure~\ref{fig4}b shows the
relation between $\Gamma$ and $\epsilon$ for these~67 stars.

We can quantify the likelihood of one scenario over the other by comparing how far each lies from the various
relations. To do so, we first performed a Bayesian linear fit to the $\epsilon$--$T_\mathrm{eff}$ and 
$\epsilon$--$\ln(\Gamma)$ relations for all stars with $T_\mathrm{eff} > 5800$ K with unambiguous identifications
(i.e. the blue points in Figures~\ref{fig4}a and b). For all stars with an uncertain identification,
we then calculated the likelihood of obtaining the observed $T_\mathrm{eff}$ for each of the two possible values of
$\epsilon$ ($\epsilon_\mathrm{A}$ and $\epsilon_\mathrm{B}$), the observed $\Gamma$, the parameters of the linear 
fits, and all respective uncertainties\footnote{Gaussian uncertainties
were assumed for $T_\mathrm{eff}$, $\epsilon$ and $\ln(\Gamma)$; uncertainties for the linear parameters were determined 
from the marginal posterior of the linear fit.}. Full details of this method will be provided in a future paper
(M. Gruberbauer et al. 2012, in prep.). The required integration over the parameter space was carried out 
using MultiNest \citep{Feroz09}. These two likelihood values, $P(T_\mathrm{eff} | \epsilon_\mathrm{A}, \Gamma)$ and 
$P(T_\mathrm{eff} | \epsilon_\mathrm{B}, \Gamma)$, were then used to calculate the Bayes factor (ratio of the likelihoods),
and hence the odds ratio and probability of each scenario, $P(\epsilon_\mathrm{A} | T_\mathrm{eff}, \Gamma)$ and 
$P(\epsilon_\mathrm{B} | T_\mathrm{eff}, \Gamma)$, assuming equal prior probability for both identifications. We
performed this calculation using both photometric and spectroscopic effective temperatures, where available. Our
preferred identification is the one with the greatest probability. We denote the value of $\epsilon$ for the 
preferred identification as $\epsilon_\mathrm{pref}$, and use $\epsilon_\mathrm{alt}$ for the alternate value.

In Table~\ref{tbl1} we list for each star its measured $\Delta\nu$, the value of $\epsilon_\mathrm{pref}$ and 
$\epsilon_\mathrm{alt}$, $\Gamma$, both effective temperatures, $T_\mathrm{eff}^\mathrm{phot}$ and 
$T_\mathrm{eff}^\mathrm{spec}$, and the probabilities, $P(\epsilon_\mathrm{pref} | T_\mathrm{eff}^\mathrm{phot}, \Gamma)$
and $P(\epsilon_\mathrm{pref} | T_\mathrm{eff}^\mathrm{spec}, \Gamma)$, of our preferred scenario. 
For most of the stars we find strong support for our preferred scenario. One star (KIC\,1725815), without a 
measured $T_\mathrm{eff}^\mathrm{spec}$, is an outlier in the $\epsilon$--$T_\mathrm{eff}$ plane. While the 
linewidth supports one scenario, $T_\mathrm{eff}^\mathrm{phot}$ supports the alternate scenario. It is not entirely 
clear which is correct,although the method of \citet{Mosser10} agrees with the alternate scenario favored 
by $T_\mathrm{eff}^\mathrm{phot}$. Spectroscopic measurement of the temperature may help resolve this case.

The $\epsilon$--$T_\mathrm{eff}$ 
relation for the final selected identification of all stars in our sample is shown in Figure~\ref{fig4}c.

\begin{deluxetable*}{lrrrccccc}
\tabletypesize{\scriptsize}
\tablecaption{Measurements of $\Delta\nu$, $\epsilon_\mathrm{pref}$, $\epsilon_\mathrm{alt}$, $\Gamma$, $T_\mathrm{eff}$ and P($\epsilon$) in 
stars with ambiguous mode identifications.\label{tbl1}}
\tablehead{
\colhead{KIC/Name} & \colhead{$\Delta\nu$} & \colhead{$\epsilon_\mathrm{pref}$} & \colhead{$\epsilon_\mathrm{alt}$} & \colhead{$\Gamma$} & \colhead{$T_\mathrm{eff}^\mathrm{phot}$} & \colhead{$T_\mathrm{eff}^\mathrm{spec}$} & \colhead{$P(\epsilon_\mathrm{pref} | \Gamma, T_\mathrm{eff}^\mathrm{phot})$} & \colhead{$P(\epsilon_\mathrm{pref} | \Gamma, T_\mathrm{eff}^\mathrm{spec})$} \\
 & \colhead{($\mu$Hz)} & & & \colhead{($\mu$Hz)} & \colhead{(K)} & \colhead{(K)} & \colhead{(\%)} & \colhead{(\%)}}
\startdata
 1430163 & 85.22 $\pm$ 0.39 & 1.08 $\pm$ 0.09 & 1.58 $\pm$ 0.06 & $4.29^{+0.54}_{-0.44}$ & 6796 $\pm$ 78 & 6520 $\pm$ 60 & 99.7 & 99.3\\
 1725815\tablenotemark{a} & 55.89 $\pm$ 0.20 & 1.51 $\pm$ 0.06 & 1.07 $\pm$ 0.06 & $1.55^{+0.08}_{-0.08}$ & 6550 $\pm$ 82 &      ---      & 73.2 & --- \\
 2837475 & 75.97 $\pm$ 0.14 & 0.84 $\pm$ 0.04 & 1.44 $\pm$ 0.06 & $9.28^{+0.69}_{-0.64}$ & 6688 $\pm$ 57 & 6700 $\pm$ 60 & 99.9 & 99.9\\
 2852862 & 53.46 $\pm$ 0.18 & 1.10 $\pm$ 0.06 & 1.55 $\pm$ 0.05 & $2.52^{+0.22}_{-0.20}$ & 6417 $\pm$ 58 &      ---      & 89.7 & --- \\
 3424541 & 41.58 $\pm$ 0.13 & 1.32 $\pm$ 0.12 & 0.81 $\pm$ 0.05 & $4.39^{+0.52}_{-0.47}$ & 6475 $\pm$ 66 & 6080 $\pm$ 60 & 73.2 & 86.7\\
 3456181 & 52.02 $\pm$ 0.15 & 1.04 $\pm$ 0.05 & 1.45 $\pm$ 0.04 & $5.01^{+0.16}_{-0.15}$ & 6732 $\pm$ 91 & 6270 $\pm$ 60 & 98.9 & 97.7\\
 3643774 & 76.15 $\pm$ 0.15 & 1.29 $\pm$ 0.06 & 0.75 $\pm$ 0.04 & $2.16^{+0.09}_{-0.09}$ & 6029 $\pm$ 104&      ---      & 100.0& --- \\
 3733735 & 92.59 $\pm$ 0.41 & 0.88 $\pm$ 0.09 & 1.43 $\pm$ 0.09 & $9.27^{+1.00}_{-0.90}$ & 6711 $\pm$ 66 & 6715 $\pm$ 60 & 99.8 & 99.8\\
 3967430 & 88.06 $\pm$ 0.25 & 0.93 $\pm$ 0.06 & 1.50 $\pm$ 0.08 & $6.10^{+0.93}_{-0.76}$ & 6622 $\pm$ 53 &      ---      & 99.4 & --- \\
 4465529 & 72.70 $\pm$ 0.23 & 1.09 $\pm$ 0.06 & 1.58 $\pm$ 0.03 & $4.42^{+0.32}_{-0.29}$ & 6345 $\pm$ 49 &      ---      & 99.7 & --- \\
 4586099 & 61.42 $\pm$ 0.22 & 1.11 $\pm$ 0.06 & 1.53 $\pm$ 0.05 & $4.61^{+0.22}_{-0.21}$ & 6383 $\pm$ 58 & 6296 $\pm$ 60 & 99.2 & 99.0\\
 4638884 & 60.46 $\pm$ 0.18 & 1.04 $\pm$ 0.06 & 1.35 $\pm$ 0.06 & $5.63^{+0.28}_{-0.27}$ & 6662 $\pm$ 57 & 6375 $\pm$ 60 & 95.9 & 93.4\\
 4931390 & 93.07 $\pm$ 0.59 & 0.79 $\pm$ 0.13 & 1.39 $\pm$ 0.04 & $5.58^{+0.79}_{-0.63}$ & 6557 $\pm$ 51 &      ---      & 92.8 & --- \\
 5431016 & 48.92 $\pm$ 0.21 & 1.21 $\pm$ 0.07 & 0.86 $\pm$ 0.07 & $2.74^{+0.14}_{-0.14}$ & 6601 $\pm$ 62 &      ---      & 93.7 & --- \\
 5516982 & 83.80 $\pm$ 0.28 & 1.14 $\pm$ 0.07 & 1.53 $\pm$ 0.06 & $2.60^{+0.13}_{-0.12}$ & 6444 $\pm$ 50 &      ---      & 89.3 & --- \\
 5773345 & 57.28 $\pm$ 0.15 & 1.08 $\pm$ 0.05 & 1.52 $\pm$ 0.04 & $3.58^{+0.13}_{-0.13}$ & 6214 $\pm$ 61 & 6130 $\pm$ 60 & 95.7 & 93.5\\
 6508366 & 51.29 $\pm$ 0.07 & 1.11 $\pm$ 0.02 & 1.48 $\pm$ 0.06 & $4.95^{+0.38}_{-0.35}$ & 6499 $\pm$ 46 & 6354 $\pm$ 60 & 98.2 & 97.7\\
 6679371 & 50.69 $\pm$ 0.16 & 0.86 $\pm$ 0.05 & 1.34 $\pm$ 0.05 & $5.07^{+0.32}_{-0.30}$ & 6598 $\pm$ 59 & 6260 $\pm$ 60 & 83.7 & 65.1\\
 7103006 & 59.34 $\pm$ 0.23 & 1.08 $\pm$ 0.07 & 1.49 $\pm$ 0.06 & $4.12^{+0.34}_{-0.32}$ & 6421 $\pm$ 51 & 6394 $\pm$ 60 & 95.7 & 96.3\\
 7206837 & 78.69 $\pm$ 0.17 & 1.17 $\pm$ 0.04 & 1.60 $\pm$ 0.05 & $4.36^{+0.43}_{-0.39}$ & 6392 $\pm$ 59 & 6304 $\pm$ 60 & 99.6 & 99.5\\
 7282890 & 45.27 $\pm$ 0.25 & 0.90 $\pm$ 0.09 & 1.43 $\pm$ 0.12 & $4.42^{+0.29}_{-0.27}$ & 6341 $\pm$ 47 & 6384 $\pm$ 60 & 74.7 & 75.4\\
 7529180 & 85.89 $\pm$ 0.28 & 0.96 $\pm$ 0.07 & 1.52 $\pm$ 0.12 & $3.27^{+0.32}_{-0.25}$ & 6682 $\pm$ 58 & 6700 $\pm$ 60 & 82.2 & 81.2\\
 7771282 & 72.55 $\pm$ 0.23 & 1.09 $\pm$ 0.06 & 1.52 $\pm$ 0.07 & $3.56^{+0.33}_{-0.27}$ & 6407 $\pm$ 74 &      ---      & 95.7 & --- \\
 7940546 & 58.67 $\pm$ 0.14 & 1.09 $\pm$ 0.04 & 1.56 $\pm$ 0.04 & $3.01^{+0.25}_{-0.22}$ & 6350 $\pm$ 111& 6264 $\pm$ 60 & 94.9 & 94.1\\
 8360349\tablenotemark{b} & 41.04 $\pm$ 0.15 & 0.97 $\pm$ 0.06 & 1.54 $\pm$ 0.09 &  ---  & 6258 $\pm$ 49 & 6340 $\pm$ 60 & 57.7 & 67.0\\
 8367710 & 55.36 $\pm$ 0.14 & 1.16 $\pm$ 0.04 & 1.54 $\pm$ 0.06 & $2.26^{+0.20}_{-0.18}$ & 6352 $\pm$ 66 & 6500 $\pm$ 60 & 83.0 & 86.7\\
 8579578 & 49.90 $\pm$ 0.15 & 1.08 $\pm$ 0.05 & 1.43 $\pm$ 0.08 & $4.62^{+0.51}_{-0.43}$ & 6308 $\pm$ 45 & 6380 $\pm$ 60 & 91.3 & 93.0\\
 9206432 & 84.51 $\pm$ 0.23 & 1.05 $\pm$ 0.06 & 1.48 $\pm$ 0.03 & $8.41^{+0.75}_{-0.69}$ & 6494 $\pm$ 46 & 6608 $\pm$ 60 & 99.9 &100.0\\
 9226926 & 73.70 $\pm$ 0.32 & 0.83 $\pm$ 0.08 & 1.14 $\pm$ 0.11 & $9.66^{+0.97}_{-0.85}$ & 7149 $\pm$ 132& 6892 $\pm$ 60 & 86.8 & 83.0\\
 9353712 & 51.37 $\pm$ 0.17 & 1.12 $\pm$ 0.06 & 1.57 $\pm$ 0.03 & $2.78^{+0.19}_{-0.17}$ & 6416 $\pm$ 56 &      ---      & 96.1 & --- \\
 9812850 & 64.59 $\pm$ 0.22 & 1.10 $\pm$ 0.06 & 1.45 $\pm$ 0.04 & $5.37^{+0.44}_{-0.40}$ & 6407 $\pm$ 47 & 6325 $\pm$ 60 & 98.1 & 98.0\\
 9908400 & 36.50 $\pm$ 0.17 & 1.00 $\pm$ 0.07 & 1.68 $\pm$ 0.06 & $2.08^{+0.17}_{-0.13}$ & 6000 $\pm$ 55 & 6400 $\pm$ 60 & 62.1 & 85.0\\
10208303 & 62.32 $\pm$ 0.37 & 0.90 $\pm$ 0.11 & 1.39 $\pm$ 0.08 &           ---          & 6665 $\pm$ 78 &      ---      & 73.8 & --- \\
10709834 & 67.98 $\pm$ 0.22 & 1.17 $\pm$ 0.06 & 0.74 $\pm$ 0.06 & $4.98^{+0.27}_{-0.25}$ & 6754 $\pm$ 56 & 6508 $\pm$ 60 & 87.7 & 92.0\\
10730618 & 66.16 $\pm$ 0.27 & 1.06 $\pm$ 0.08 & 1.53 $\pm$ 0.08 & $2.78^{+0.11}_{-0.10}$ & 6422 $\pm$ 54 &      ---      & 85.5 & --- \\
10909629 & 49.81 $\pm$ 0.27 & 0.94 $\pm$ 0.09 & 1.47 $\pm$ 0.06 & $3.39^{+0.36}_{-0.33}$ & 6501 $\pm$ 61 &      ---      & 82.0 & --- \\
11081729 & 90.03 $\pm$ 0.20 & 1.03 $\pm$ 0.04 & 1.48 $\pm$ 0.06 & $6.60^{+0.72}_{-0.65}$ & 6605 $\pm$ 51 & 6630 $\pm$ 60 & 99.7 & 99.8\\
11128126 & 77.36 $\pm$ 0.28 & 1.11 $\pm$ 0.12 & 0.73 $\pm$ 0.07 &           ---          & 6250 $\pm$ 55 &      ---      & 73.2 & --- \\
11253226 & 77.30 $\pm$ 0.20 & 0.81 $\pm$ 0.05 & 1.34 $\pm$ 0.04 & $8.72^{+0.50}_{-0.48}$ & 6682 $\pm$ 51 & 6605 $\pm$ 60 & 99.5 & 99.4\\
11290197\tablenotemark{b} & 74.56 $\pm$ 0.47 & 0.85 $\pm$ 0.12 & 1.29 $\pm$ 0.17 &  ---  & 6544 $\pm$ 63 &      ---      & 58.0 & --- \\
11401708 & 40.00 $\pm$ 0.14 & 1.31 $\pm$ 0.07 & 0.70 $\pm$ 0.06 & $1.35^{+0.20}_{-0.16}$ & 5859 $\pm$ 63 &      ---      & 100.0& --- \\
12069127 & 48.47 $\pm$ 0.13 & 1.01 $\pm$ 0.04 & 1.49 $\pm$ 0.02 & $2.66^{+0.12}_{-0.11}$ & 6476 $\pm$ 66 &      ---      & 77.3 & --- \\
12555505 &108.08 $\pm$ 0.61 & 1.52 $\pm$ 0.12 & 0.84 $\pm$ 0.12 &           ---          & 5704 $\pm$ 66 &      ---      & 91.8 & --- \\
Procyon  & 56.20 $\pm$ 0.35 & 1.06 $\pm$ 0.11 & 1.58 $\pm$ 0.08 & $2.86^{+1.75}_{-0.85}$ &      ---      & 6530 $\pm$ 50 & ---  & 93.4\\
HD 49933 & 85.53 $\pm$ 0.18 & 1.06 $\pm$ 0.04 & 1.54 $\pm$ 0.05 & $6.57^{+1.09}_{-0.98}$ &      ---      & 6750 $\pm$ 130& ---  & 99.9\\
HD 181420& 75.20 $\pm$ 0.32 & 0.92 $\pm$ 0.09 & 1.36 $\pm$ 0.07 & $7.65^{+1.30}_{-1.11}$ &      ---      & 6580 $\pm$ 105& ---  & 97.5
\enddata
\tablenotetext{a}{Identification favored by $T_\mathrm{eff}^\mathrm{phot}$ disagrees with identification favored by $\Gamma$.}
\tablenotetext{b}{Identification made with the aid of $\Delta\nu$.}
\end{deluxetable*}

\section{Conclusions}
We have presented a method to effectively determine the correct mode identification in stars for which this has previously
been a problem. These are the F stars with large linewidths that make it difficult to distinguish the $l=0$
and $l=2$ modes. This method uses the relationship between effective temperature, mode linewidth and $\epsilon$ to 
determine what values of $\epsilon$ are reasonable for the star, and therefore which of the two possible scenarios 
is most likely correct. This method provides robust results in the vast majority of cases because the value of $\epsilon$ implied 
by each scenario is very distinct, even in low signal-to-noise stars, representing a major improvement over previous 
methods. For the few cases that are still ambiguous, additional information, such as the large separation, $\Delta\nu$, 
can be included to help resolve the matter. 

\acknowledgments
The authors gratefully acknowledge the {\it Kepler} Science Team and all those who have contributed to the {\it Kepler Mission} 
for their tireless efforts which have made these results possible. Funding for the {\it Kepler Mission} is provided by 
NASA’s Science Mission Directorate.
We acknowledge the support of the Australian Research Council. TRW is supported by an Australian Postgraduate Award, a
University of Sydney Merit Award, an Australian Astronomical Observatory PhD Scholarship and a Denison Merit Award. 
MG received financial support from an NSERC Vanier Scholarship. SH acknowledges financial support from the
Netherlands Organisation of Scientific Research (NWO).

\end{document}